\pdfoutput=1
\documentclass[conference]{IEEEtran}
\IEEEoverridecommandlockouts
\usepackage{cite}
\usepackage{amsmath,amssymb,amsfonts}
\usepackage{algorithmic}
\usepackage{graphicx}
\usepackage{textcomp}
\usepackage{xcolor}
\usepackage{comment}
\usepackage{balance}
\usepackage[hidelinks]{hyperref}
\def\BibTeX{{\rm B\kern-.05em{\sc i\kern-.025em b}\kern-.08em
    T\kern-.1667em\lower.7ex\hbox{E}\kern-.125emX}}
\begin{document}

\IEEEpubid{\makebox[\columnwidth]{Invited paper for 2023 IEEE International Conference on eScience \hfill} \hspace{\columnsep}\makebox[\columnwidth]{ }}

\title{Research Software Engineering in 2030
\thanks{DSK was partially supported by Alfred P.\ Sloan Foundation award 10073. SH is supported by grant EP/S021779/1.}
}

\author{\IEEEauthorblockN{Daniel S. Katz}
\IEEEauthorblockA{\textit{NCSA \& CS \& ECE \& iSchool} \\
\textit{University of Illinois Urbana Champaign}\\
Urbana, IL, USA \\
ORCID: 0000-0001-5934-7525}
\and
\IEEEauthorblockN{Simon Hettrick}
\IEEEauthorblockA{\textit{Software Sustainability Institute} \\
\textit{University of Southampton}\\
Southampton, UK \\
ORCID: 0000-0002-6809-5195}
}

\maketitle

\begin{abstract}
This position paper for an invited talk on the ``Future of eScience'' discusses the Research Software Engineering Movement and where it might be in 2030. Because of the authors' experiences, it is aimed globally but with examples that focus on the United States and United Kingdom.
\end{abstract}


\section{Introduction}

Research software has existed since at least the early 1950s. Formal software engineering started in the 1960s to understand operational software such as operating systems and NASA flights, and later business software, with research software infrequently the focus of study. Bringing these two concepts together  and identifying this as a named discipline---research software engineering (RSEng)---with a complementary set of practitioners---Research Software Engineers (RSEs)---was first formalized by a small group in the UK in 2012, as described by Brett et al. in 2017~\cite{Brett2017}.

Research software is essential to most research~\cite{Hettrick2014} and research software engineering (building and maintaining software in a research context using relevant software engineering knowledge) is performed by many people, including RSEs who focus on this work, as well as others in the research community.
The RSE Community has grown substantially since 2012. Today (August 2023), there are eight national and multinational RSE associations with established communities and activities, and at least two more developing, all coordinating via an International Council of RSE Associations~\cite{RSEcouncil}.
The largest, the UK-based Society of Research Software Engineering, has a community of about 4800 people, including about 700 dues-paying members, while the next largest, the US Research Software Association, has about 1900 members. At least 40 UK universities have formal RSE groups, as do many US universities and national laboratories.

In 2021, Cohen et al.~\cite{Cohen2021} defined  research software engineering as being built on four pillars: software development, community, training, and policy.  In the rest of this position paper, we use these pillars and hypothesize about how the RSE landscape will have changed in 2030, an intervening period that represents a little over half of the current age of the RSE community.

\section{How Research Software Engineering might be in 2030}

In this section, we predict some qualitative changes to research software engineering that we might expect by 2030. Where it seems reasonable to us, we also make quantitative predictions that we consider plausible, and perhaps even probable.

\textbf{Software development}:

\begin{itemize}
\item Research software developers (including but not limited to RSEs) will be generally aware of research software engineering as a concept and its best practices, at about the same level that software developers today are aware of source control packages like git. This will be due to a greater awareness of the role of software in research and changes to community, training, and policy.
\item This greater awareness of research software, and the existence and availability of RSEs, will lead to research software being more reliable, maintainable, and better tested. This change to one of the most important underpinning tools of research will make research results more reliable and reproducible.
\item More research software will be sustained because funders will better understand that funding the maintenance of research software will improve the return on their initial investment into it. This switch to recognising and supporting successful research software will increase awareness of successful research software and reduce reinvention of existing functionality.
\item Research software developers will begin to receive recognition, not just for their contribution to research, but also for their contribution to software within academia. This will allow the design of an incentive system that aligns career progression for research software developers with the practices needed to develop reliable and reproducible research software.
\end{itemize}

\textbf{Community}:

\begin{itemize}
\item At least 50 countries globally will have RSE associations.
\item While RSEs will generally be members of their own local RSE organization, there will also be communication channels they use to talk to and collaborate with RSEs in other national organizations.
\item There will be a global flagship RSE conference that moves to a different country each year and that also supports virtual participation.
\item There will be at least one research software engineering journal.
\item There will be at least 50 professors whose focus is on research software engineering.
\item Most professional scholarly societies will have research software tracks or themes in their conferences, and will offer prizes to research software developers.
\item Awareness of the RSE career will be significantly more widespread, which will lead to around a quarter of new recruits being recruited after their undergraduate degree.
\end{itemize}

\textbf{Training}:

\begin{itemize}
\item At least one university Masters of Research Software Engineering (MRSE) program will exist, and many universities will offer RSE certificates or minors at the undergraduate level.
\item Some secondary students will be aware of research software, and will consider an RSE career when entering university.
\item Both Software Carpentry and a merged program beyond it, similar to programs today such as INTERSECT~\cite{intersect} and CodeRefinery~\cite{CodeRefinery}, will offer frequent synchronous and asynchronous research software engineering training.
\item As a part of more standardized RSE career paths, skills required at various levels will be specified and training for RSEs new to these levels will be available both synchronously and asynchronously.
\item Training for the initial RSE level will be offered in internship programs.
\item Courses in software engineering techniques will be either mandatory or strongly encouraged for all new doctoral students.
\end{itemize}

\textbf{Policy}:

\balance

\begin{itemize}
\item The value of software, and its vital role in the production of research results, will be accepted across all levels in research-performing organizations and by policymakers across government and funders.
\item In the US, 75\% of R1 universities will have RSE organizations. In the UK, all research-led universities will host an RSE Group.
\item Funding organizations will have specific programs to fund 1) the development of new sustainable research software, 2) the maintenance and sustaining of existing research software, and 3) the transition of software developed in research activities into sustainable research software.
\item Most US universities will have RSE career paths that contain multiple levels, allow for both technical and management career progression over 30+ years, and allow RSEs to be PIs.
\item RSE career paths will be mostly standardized, and it will be as easy to move from one organization to another at a given RSE level as it as today at a given faculty level.
\item Institutions with RSE groups will also have Open Source Program Offices and vice-versa, and they will work together to promote and strengthen research software at their institutions directly and via policy changes.
\item Research-performing organizations (e.g., universities, national laboratories) will recognize and reward research software work in all research-performing positions, as will national research assessment activities.
\end{itemize}

\section{The value of predictions}

While the predictions we make in this position paper may not come to pass, as we've said, we think they are plausible. In some cases, they simply extend current trends, while in others, they require new collaborative work to accomplish. We urge community members to consider how they might make these predictions more like to be fulfilled, working within existing RSE communities and with organizations such as the Software Sustainability Institute~\cite{SSI}, the US Research Software Sustainability Institute~\cite{URSSI}, and the Research Software Alliance~\cite{ReSA}.

\section*{Acknowledgment}

The authors thank all members of the international RSE community for their collective work in getting us to where we are now, defining a vision for the future, and moving us all towards it.

\bibliographystyle{IEEEtransDOI}
\bibliography{references}

\end{document}